\documentclass[sigconf,10pt]{acmart}
\AtBeginDocument{%
  }

\renewcommand\footnotetextcopyrightpermission[1]{} % removes footnote with conference information in first column
\setcopyright{none}
\authorsaddresses{}

\usepackage{pifont}
\usepackage{algorithm}
\usepackage[noend]{algpseudocode} 
\usepackage{multirow}
\usepackage{xspace}
\usepackage{xcolor}
\usepackage[font=small, skip=0pt]{caption}
\usepackage[font=small, skip=0pt]{subcaption}
\usepackage{tabularray}
\usepackage{svg}
\usepackage{tabularx}
\usepackage{graphicx}
\usepackage{changepage} 
\usepackage{enumitem}
\usepackage{tabularx}
\usepackage{booktabs}
\usepackage{subcaption} 
\usepackage{placeins}
\def\name{Dora\xspace}

\newcommand{\objfunc}[1]{$|T_{plan latency} - T_{SLO latency}|/ (E_{global})^{\alpha}$}

\definecolor{cardinal}{rgb}{0.77, 0.12, 0.23}
\newcommand\mycommfont[1]{\small\textcolor{cardinal}{\textrm{#1}}}
\expandafter\def\expandafter\normalsize\expandafter{%
    \normalsize%
    \setlength\abovedisplayskip{0pt}%
    \setlength\belowdisplayskip{0pt}%
    \setlength\abovedisplayshortskip{-10pt}%
    \setlength\belowdisplayshortskip{0pt}%
}
\newenvironment{denseitemize}{
\begin{itemize}[topsep=2.5pt, partopsep=0pt, leftmargin=1.5em]
  \setlength{\itemsep}{2.5pt}
  \setlength{\parskip}{0pt}
  \setlength{\parsep}{0pt}
}{\end{itemize}}

\usepackage{titlesec}
\titleformat{\paragraph}
  [runin]               % run-in style (not on a new line)
  {\bfseries}           % bold font
  {}                    % label (empty means no numbering)
  {0pt}                 % spacing between label and title
  {}                    % code before the title

\def\ie{{i.e.}}
\def\eg{{e.g.}}

\titlespacing*{\paragraph}
{0pt}{1ex plus 0.5ex minus 0.2ex}{0.5em}  

\usepackage{tikz}

\newcommand*\blackcircled[1]{\tikz[baseline=(char.base)]{
            \node[shape=circle,fill,inner sep=1pt] (char) {\textcolor{white}{#1}};}}
            
\begin{document}

\title{Dora: QoE-Aware Hybrid Parallelism for Distributed Edge AI}

% \author{Anonymous Authors}

% \email{john.doe@example.com}
% %% \orcid{1234-5678-9012}
% %% \author{G.K.M. Tobin}
% %% \authornotemark[1]
% %% \email{webmaster@marysville-ohio.com}
% \affiliation{%
%   \institution{Somewhere}
%   \city{Some city}
%   \state{Some state}
%   \country{Some country}
%   \postcode{12345-6789}
% }

\author{Jianli Jin}
\authornotemark[1]
\affiliation{%
  \institution{UIUC}
  \country{}
  }
\email{jianlij2@illinois.edu}

\author{Ziyang Lin}
\affiliation{%
  \institution{UIUC}
  \country{}
  }
\email{ziyang10@illinois.edu}
\authornote{Both authors contributed equally to this research.}

\author{Qianli Dong}
\affiliation{%
  \institution{Northwestern University}
  \country{}
  }
\email{qianlid@northwestern.edu}

\author{Yi Chen}
\affiliation{%
  \institution{University of California, Riverside}
  \country{}
  }
\email{ychen1329@ucr.edu}

\author{Jayanth Srinivasa}
\affiliation{%
  \institution{Cisco Research}
  \country{}
  }
\email{jasriniv@cisco.com}

\author{Myungjin Lee}
\affiliation{%
  \institution{Cisco Research}
  \country{}
  }
\email{myungjle@cisco.com}

\author{Zhaowei Tan}
\affiliation{%
  \institution{University of California, Riverside}
  \country{}
  }
\email{ztan@ucr.edu}

\author{Fan Lai}
\affiliation{%
  \institution{UIUC}
  \country{}
  }
\email{fanlai@illinois.edu}

\renewcommand{\shortauthors}{Jin et al.}

\begin{abstract}
  \label{sec:abstract}
With the proliferation of edge AI applications, satisfying user quality of experience (QoE) requirements, such as model inference latency, has become a first-class objective, as these models operate in resource-constrained settings and directly interact with  users. Yet, modern AI models routinely exceed the resource capacity of individual devices, necessitating distributed execution across heterogeneous devices over variable and contention-prone networks. Existing planners for hybrid (e.g., data and pipeline) parallelism largely optimize for throughput or device utilization, overlooking QoE, leading to severe resource inefficiency (e.g., unnecessary energy drain), or QoE violations under runtime dynamics.

We present Dora, a framework for QoE-aware hybrid parallelism in distributed edge AI training and inference. Dora jointly optimizes heterogeneous computation, contention-prone networks, and multi-dimensional QoE objectives via three key mechanisms: (i) a heterogeneity-aware model partitioner that determines and assigns model partitions across devices, forming a compact set of QoE-compliant plans; (ii) a contention-aware network scheduler further refines these candidate plans by maximizing compute–communication overlap; and (iii) a runtime adapter that adaptively composes multiple plans to maximize global efficiency while respecting overall QoEs. Across representative edge deployments—including smart homes, traffic analytics, and small edge clusters—Dora achieves 1.1--6.3$\times$ faster execution and, alternatively, reduces energy consumption by 21–82\%, all while maintaining QoE under runtime dynamics.
\end{abstract}

\maketitle
\pagestyle{plain}

\section{Introduction}

The proliferation of on-device AI applications---ranging from large language models (LLMs)-powered assistants~\cite{Wang_2025, yu2024edgellmenablingefficientlarge, li2024personalllmagentsinsights} and smart robotics~\cite{wang2024largelanguagemodelsrobotics, brohan2023rt2visionlanguageactionmodelstransfer} to video surveillance~\cite{desilva2025largelanguagemodelsvideo} with multimodal LLMs---has brought Quality of Experience (QoE) to the forefront of edge AI tuning and inference.
Unlike cloud-based deployments, where the system's primary objective is to maximize throughput~\cite{zhong2024distservedisaggregatingprefilldecoding, Wang_2025} or resource efficiency~\cite{298679} for aggregate performance across users, on-device AI often serves individual users, needing to satisfy per-user QoE requirements such as training or serving latency~\cite{10.1145/3093337.3037698,10.1145/3241539.3241559} and energy efficiency~\cite{Wang_2025}.

Aligning execution with these QoE targets not only ensures service quality but also improves system efficiency: exceeding QoE thresholds yields diminishing perceptual returns while wasting scarce resources. For example, in federated learning, clients only need to train quickly enough to avoid delaying the global aggregation round across hundreds of clients~\cite{lai2022fedscalebenchmarkingmodelperformance,273723}; in continuous or personalized learning, training needs only keep pace with data generation (e.g., sensor events or app usage), thereby reducing interference with foreground tasks and conserving energy. On the inference side, users interacting with on-device LLMs expect token generation rates aligned with human reading or interaction speed~\cite{andes-arxiv24, tempo-arxiv25}, and edge cameras for traffic analytics need only track the natural rate of scene evolution~\cite{aifarming-2025, assistai-asset25}.

However, meeting QoE requirements on a single edge device is increasingly untenable (\S\ref{subsec:qoe-background}). Emerging applications depend on high-accuracy models that, while modest by cloud standards, saturate the compute, memory, and energy envelopes of edge devices. Running Qwen3-4B alone requires roughly 12~GB of memory, while iPhone~15 offers only 6GB of RAM. Worse, even a fully charged iPhone~15 can sustain less than an hour of operation when running assistive edge-AI agents (e.g., small multimodal LLMs for visually impaired individuals)~\cite{mobilellm-icml24, assistai-asset25}. 
Increasingly, edge environments have multiple compute-capable devices---smartphones, tablets, laptops in a home, or multiple AI cameras at an intersection---creating opportunities for QoE-aware, distributed model execution across edge devices. The goal is not merely to eliminate resource slack or maximize raw throughput, but to shape execution around user-driven QoE objectives while operating under the dynamism of real-world edge environments.

Enforcing QoE-aware hybrid parallelism, such as combining data and pipeline parallelism across edge devices, introduces unique challenges (\S\ref{sec:background}). Unlike controllable, homogeneous cloud settings with abundant and predictable network bandwidth, edge deployments consist of devices with heterogeneous compute and memory capabilities, and communication over contention-prone wireless channels (e.g., WiFi) or irregular wired structures (e.g., chains or rings). 
Even very recent edge AI advances for distributed model execution~\cite{ye_2024_asteroid} have largely overlooked these network effects, leading to 4$\times$ efficiency degradation, as contention erases the benefits of model parallelism. Under dynamics, these systems frequently trigger heavy-weight replanning or model migration, stalling the service, while their emphasis on maximizing raw throughput can cascade into QoE violations such as rapid battery drain or thermal throttling.

This paper introduces \name, a framework for distributed edge AI training and inference that generates parallelism plans explicitly to satisfy user-defined QoE objectives (e.g., execution latency, energy usage). By aligning application-level performance needs with underlying model execution, \name identifies the optimal balance between QoE and system efficiency, preventing unnecessary resource consumption while supporting existing edge AI software stacks with a few lines of code changes in APIs (\S\ref{sec:overview}).

Overcoming the aforementioned challenges requires management over the interplay of heterogeneity, contention, and multi-objective requirements. Even without network effects or QoE constraints, identifying an optimal hybrid parallelism plan under heterogeneous compute resources is already NP-hard~\cite{metis-atc24}. Network contention further exacerbates this challenge: computation dependencies that span shared links and irregular topologies can negate the gains of distributed computation, rendering compute-optimized plans ineffective. \name tackles this interplay through: (i) a model partitioner that accounts for device heterogeneity to identify a compact set of QoE-compliant, compute–energy efficient candidate plans (\S\ref{sec:model-partitioner}); and (ii) a contention-aware network scheduler that refines these candidates by adaptively allocating communication over networks to maximize computation–communication overlap, enabling \name to select the plan that maximizes energy efficiency while satisfying user-specified QoE objectives (\S\ref{sec:network-scheduler}).

Yet (offline) planning alone is insufficient, as edge environments are inherently dynamic: device capabilities shift due to background workloads or thermal events, and bandwidth fluctuates on subsecond timescales, rendering a pre-determined plan quickly suboptimal or even infeasible. To sustain QoE, \name employs a runtime adapter that composes and switches among the Pareto-optimal plans over time, maximizing global efficiency while satisfying long-term QoE targets. It coordinates with the network planner to deliver rapid, low-overhead network re-planning to absorb transient dynamics, while minimizing the cost of plan switching. This design enables second-scale responsiveness to dynamics while maintaining high overall efficiency (\S\ref{sec:runtime-adapter}).

We evaluate \name across representative real-world edge AI deployments—including smart homes, traffic monitoring systems, and small edge clusters—and across modern model families (\eg, Qwen3) for both training and inference workloads. Our results show that \name delivers 1.1--6.3$\times$ higher model execution speed compared to state-of-the-art hybrid parallelism planners~\cite{ye_2024_asteroid, metis-atc24, megatron-sc21}, and, to the best of our knowledge, represents the first framework to enable QoE-aware hybrid parallelism for edge AI. While meeting target QoE levels, \name achieves 20.7--82\% resource savings and adapts to runtime dynamics within seconds, ensuring sustained and efficient edge AI service.

Overall, this paper makes the following contributions:

\begin{itemize}

\item We introduce a new QoE-driven formulation of distributed model parallelism for edge AI to maximize resource efficiency without hurting user experience.

\item We design a novel hybrid-parallelism planner to account for heterogeneous compute, network contention, and dynamics, thereby maximizing QoE and efficiency.

\item We implement \name and demonstrate 1.1--6.3$\times$ efficiency gains across real-world edge AI settings.

\end{itemize}
% \lee{the sentence in 2nd bullet point seems to be terse by omitting some necessary words; e.g., (thereby) maximizing. for 3rd bullet point, "its" doesn't seem to be necessary.}\fan{thanks}

\section{Background and Motivation}
\label{sec:background}

We begin with a primer on QoE in edge AI, followed by the challenges it faces (\S\ref{subsec:qoe-background}), and conclude with the limitations of current advances that motivate our work (\S\ref{subsec:limitation}).

\subsection{Quality of Experience in Edge AI}
\label{subsec:qoe-background}

Edge AI has become essential as end-user applications increasingly require real-time, privacy-preserving, and context-specific intelligence that cloud execution cannot reliably or cost-effectively provide~\cite{oort-osdi21, mobilellm-icml24}. Tasks---such as on-device model tuning and personalization (e.g., federated learning, continual adaptation to user-specific behavior), as well as latency-sensitive inference for assistants, robotics, and emerging on-device AI agents---require executing models in situ, close to the user and environment.

\paragraph{QoE Requirements in  Edge AI.}
For practical use, model execution must satisfy user-specific QoE constraints (e.g., on latency, compute, memory, and energy). For example, model personalization should keep pace with data generation or complete within natural idle windows (e.g., charging periods or user sleep)~\cite{oort-osdi21, fedscale-icml22}; LLM generation needs to outpace user reading speed to avoid interaction stalls (\eg, < 200 ms per token~\cite{andes-arxiv24})\footnote{LLMs process and generate text in units of tokens. For instance, the word “streaming” may be broken down into two tokens: “stream” and “ing.”}; AR Navigation should process frames in 50 ms~\cite{ar-mobisys24}; and all executions must remain within device-level energy budgets~\cite{aifarming-2025, mobilellm-icml24}. 

\begin{figure}[t]
    \centering
    \begin{subfigure}{0.48\linewidth}
        \captionsetup{labelfont=normalfont,textfont=normalfont}
        \centering
        \includegraphics[width=\linewidth]{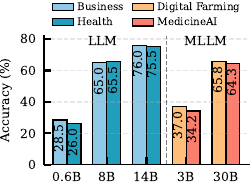}
        \caption{Qwen-3/Omni Performance.}
        \label{fig:qwen3-text}
    \end{subfigure}
    \hfill
    \begin{subfigure}{0.48\linewidth}
         \captionsetup{labelfont=normalfont,textfont=normalfont}
         \centering
         \includegraphics[width=\linewidth]{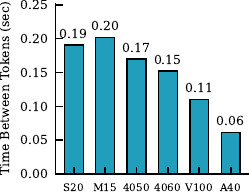}
         \caption{Qwen3-0.6B speed.}
         \label{fig:qwen-speed}
     \end{subfigure}

    \caption{(a) Edge AI desires higher-accuracy (thus often larger) models to unlock more applications, which yet often exceed the capacity of single devices. (b) Increasingly, edge environments have devices with comparable compute.}
    \label{fig:sizeaccuracy_LLM}
\end{figure}

Achieving practical use further requires high model accuracy (e.g., >75\% for on-device health assistants), which pushes developers toward using larger models: our evaluation (Figure~\ref{fig:qwen3-text})  of modern AI models (e.g., Qwen3 and Qwen2.5-Omni) on MMLU and MMMU-Pro benchmarks~\cite{mmlu-arxiv} shows about 50\% accuracy gains when scaling from 0.6B to 14B parameters. Yet such models often exceed typical device envelopes. Even an INT4-quantized Qwen3-14B still demands >8 GB RAM for inference, whereas devices like the iPhone 15 Plus offer only 6 GB, resulting in the fundamental tension between QoE requirements and on-device capability.

% \fan{"the standard iPhone 15 and 15 Plus have 6GB RAM; Nvidia Jetson Orin Nano has 8 GB RAM" "e.g. a user’s phone+tablet+laptop might collectively have 20+ GB RAM and many TOPS of compute"}

At the same time, we notice that edge environments increasingly include multiple compute-capable devices (Figure~\ref{fig:qwen-speed}), such as smartphones (\eg, Samsung S20), tablets, and laptops (\eg, NVIDIA RTX 4060) at home or networks of cameras in traffic monitoring, opening a new opportunity for QoE-aware, distributed model training and inference across multiple devices to adapt to user QoE requirements.

% \begin{table}[t]
% \centering
% \begin{tabular}{lc}
% \toprule
% \textbf{Application} & \textbf{Latency} \\
% \midrule
% AR Navigation & $<50$\,ms \\
% On-device LLM Chat & $<200$\,ms/token \\
% Federated Learning & Flexible \\
% Personalization (e.g., keyboards) & $<500$\,ms \\
% Smart Cameras & $<33$\,ms/frame \\
% \bottomrule
% \end{tabular}
% \caption{Latency requirements of common edge AI applications. \fan{better to have tuning ones.}}
% \label{tab:qoe-examples}
% \end{table}

\subsection{Limitations of Existing Solutions}
\label{subsec:limitation}

Existing advances in distributed model execution, such as Megatron-LM~\cite{megatron-sc21} and PyTorch-FSDP~\cite{fsdp-vldb23}, primarily rely on three forms of \emph{hybrid parallelism} that partition model (e.g., weights) or data across machines for collaborative execution: 
\begin{denseitemize}
\item \emph{Data Parallelism}, which partitions the dataset across machine groups, each maintaining a full model replica and synchronizing updates after every training iteration; 

\item \emph{Pipeline Parallelism}, which splits the model into sequential layers assigned to different machines, passing intermediate layer outputs between stages; 

\item \emph{Tensor Parallelism}, which partitions individual layers into sub-tensors across machines to further reduce per-machine memory footprint.
\end{denseitemize}

While effective in cloud-scale environments, these approaches, and even recent edge-targeted systems such as Asteroid~\cite{ye_2024_asteroid}, face fundamental limitations in meeting QoE requirements in resource-constrained and heterogeneous edge environments under dynamics (\eg, bandwidth changes).

% To enable efficient distributed model execution, existing advances, such as Megatron-LM~\cite{megatron-sc21} and PyTorch FSDP~\cite{fsdp-vldb23}, rely on three major forms of model parallelism that decides how the model state (\eg, weights) are partitioned across machines for collective : \lee{we never formally defined model parallelism. So, I am assuming pipeline or tensor parallelisms are part of model parallelism. My understanding is that it's a generic term to describe techniques that partition a model and run it across GPUs. Otherwise, we have to define it precisely. It may cause confusion.} 

\paragraph{L1: Hardware and network heterogeneity and contention degrade efficiency.}
Edge devices vary widely in compute and memory capacity, and their inter-device communication often occurs over shared, irregular, and contention-prone links (e.g., WiFi meshes or chained camera topologies). However, existing distributed execution frameworks commonly assume homogeneous hardware~\cite{280874} and/or uniform, contention-free D2D bandwidth (e.g., Asteroid~\cite{ye_2024_asteroid}). In real edge environments, these assumptions routinely break down, causing parallelism plans to become suboptimal or invalid and leading to substantial efficiency losses.

\begin{figure}[t]
    \centering
    \begin{subfigure}{0.49\linewidth}
        \captionsetup{labelfont=normalfont,textfont=normalfont}
        \centering
        \includegraphics[width=\linewidth]{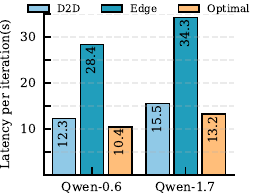}
        \caption{Training Speed.}
    \end{subfigure}
     \hfill
     \begin{subfigure}{0.49\linewidth}
         \captionsetup{labelfont=normalfont,textfont=normalfont}
         \centering
         \includegraphics[width=\linewidth]{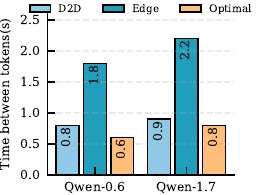}
         \caption{Inference Speed.}
     \end{subfigure}

    \caption{Existing advances overlook heterogeneous hardware and network contention, leading to poor efficiency.}
    \label{fig:heteroD_computability}
\end{figure}

As shown in Figure~\ref{fig:heteroD_computability}, we evaluate a home deployment with two laptops and two phones connected via a shared 650 Mbps WiFi network using Asteroid. We compare three settings: (1) \emph{D2D}, an idealized environment where every device pair is given a dedicated 650 Mbps link with no contention; (2) \emph{Edge}, where Asteroid's generated parallelism plan is executed on the actual WiFi network; and (3) \emph{Optimal}, a brute-force search that identifies the true optimal plan under the real network conditions. We observe that Asteroid's plan suffers a 2.4$\times$ latency degradation when deployed in the real environment due to mismatches between assumed and actual network behavior, and remains 2.8$\times$ slower than the optimal hybrid parallelism plan.

\begin{figure}[t]
    \centering

    \begin{subfigure}[t]{0.46\linewidth}
        \captionsetup{labelfont=normalfont,textfont=normalfont}
        \centering
        \includegraphics[width=\linewidth]{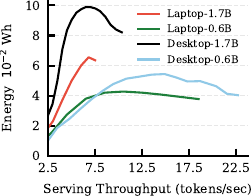}
        \caption{Energy vs. Inference Speed.}
        \label{fig:energy-speed}
    \end{subfigure}
    \hfill
    \begin{subfigure}[t]{0.46\linewidth}
        \captionsetup{labelfont=normalfont,textfont=normalfont}
        \centering
        \includegraphics[width=\linewidth]{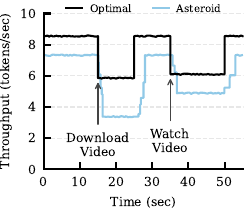}
        \caption{Serving under dynamics.}
        \label{fig:fluctuation:speed-time}
    \end{subfigure}
    \caption{Existing advances ignore QoE, incurring ineffective efficiency and failing to sustain service under dynamics.}
    \label{fig:energy}
\end{figure}

\paragraph{L2: Ignoring QoE constraints leads to ineffective execution.}
Beyond suboptimal execution speed, existing hybrid parallelism planners largely ignore users' QoE and device constraints, resulting in ineffective execution and excessive resource consumption.  
As shown in Figure~\ref{fig:energy-speed}, generating the same response with Qwen-3 at different target speeds leads to starkly different energy costs: modestly slowing generation, while still meeting QoE, can reduce energy usage by more than an order of magnitude. Energy consumption further varies across devices (e.g., Laptop vs.\ Desktop by 2.5$\times$), indicating that aligning execution with QoE targets and device characteristics with informed parallelism can unlock substantial efficiency gains.

\paragraph{L3: Slow responsiveness fails to handle runtime dynamics.}
Maintaining QoE in edge environments requires rapid adaptation: devices may start foreground applications, experience bandwidth fluctuations, or enter thermal throttling, all of which can invalidate existing parallelism plans, spike latency, and stall service (e.g., due to plan switching).

Figure~\ref{fig:fluctuation:speed-time} shows a representative case where a user downloads and streams a video, introducing both network and compute interference. These common dynamics increase the end-to-end distributed inference latency of the Asteroid-generated plan by more than 38\%. Unfortunately, existing planners require minutes to replan and switch plans (\S\ref{eval:ablation}), orders of magnitude slower than typical edge fluctuations and rendering QoE guarantees infeasible under dynamics.

\section{\name Overview}
\label{sec:overview}

In this section, we introduce \name, a QoE-aware hybrid parallelism planner for distributed ML model execution in edge environments, including training, fine-tuning, and inference.

\paragraph{Design Space.}
\name accounts for heterogeneous device capabilities (e.g., compute capability, network topology) and runtime dynamics to meet per-user QoE requirements such as execution speed and energy usage. This makes \name applicable to diverse modern edge settings, such as smart homes with multiple personal devices, LAN-connected sensor networks (e.g., camera clusters), and edge clusters with a number of servers. We formalize QoE-aware hybrid parallelism planning in \name as the following optimization problem:

\begin{align}
\label{theobjfunction-con}
\min \quad & \sum_{i \in \mathcal{D}} E_{\text{plan}}^{i} 
     \\
\text{s.t.} \quad 
& T_{\text{plan}} \leq T_{\text{QoE}}
    && \text{(E2E latency constraint)} \nonumber \\
& E_{\text{plan}}^{i} \leq E_{\text{QoE}}^{i},\ \forall i \in \mathcal{D}
    && \text{(Device energy constraint)} \nonumber \\
& M_{\text{plan}}^{i} \leq M_{\text{QoE}}^{i},\ \forall i \in \mathcal{D}
    && \text{(Device memory constraint)} \nonumber
\end{align}
\vspace{.05cm}

Here, $T_{\text{plan}}$ denotes the model execution latency under a given parallelism plan, while $T_{\text{QoE}}$ specifies the latency requirement (\eg, < 200 ms/token). The planner must respect each device's local energy and memory budgets $(E_{\text{QoE}}^{i}, M_{\text{QoE}}^{i})$. This formulation accommodates additional QoE constraints common in edge deployments (e.g., limiting computational usage to avoid interfering with foreground apps) without altering the core design.

However, directly solving this constrained problem is challenging due to the intertwined effects of heterogeneous compute capabilities, contention-prone networks, and multi-dimensional QoE requirements. Even the \emph{subproblem} of finding an optimal parallelism plan under heterogeneous compute \emph{alone} is NP-hard~\cite{metis-atc24}. Worse yet, even when the computation graph is fixed, variations in inter-device communication patterns and network contention can lead to large swings in execution latency (\S\ref{subsec:limitation}).

To make this space tractable while retaining QoE fidelity, \name applies a Lagrangian relaxation~\cite{lag-lp} to convert the latency constraint into an unconstrained objective:
\begin{align}
\min\ \sum_{i \in \mathcal{D}} E_{\text{plan}}^{i}
\;+\;
\lambda \cdot \bigl(T_{\text{plan}} - T_{\text{QoE}}\bigr)_{+} \label{theobjfunction}
\end{align}

where $(\cdot)_{+}$ penalizes only QoE violations and $\lambda$ controls the tradeoff between energy savings and latency slack. 

% \lee{can the device energy constraints in eq(1) result in a case where there is no feasible solution?}
% \zt{Since previous paragraph mentioned conditions like network structure, I would love to see which notations here are related. Not necessarily need to elaborate but some hints with pointers would be good. Similarly, how's the characteristics of ML task itself modeled (or does it affect the formulation here?}

% \begin{align}
% \label{theobjfunction}
% \min \quad & \frac{|T_{\text{plan}} - T_{\text{QoE}}|}{\left(\sum_{i \in \mathcal{D}} E_{\text{plan}}^{i}\right)^{\alpha}} \label{eq:objective}\\
% \text{s.t.} \quad 
% & E_{\text{plan}}^{i} \leq E_{\text{QoE}}^{i}, \ \forall i \in \mathcal{D} && \text{(Energy constraint)} \nonumber \\
% & M_{\text{plan}}^{i} \leq M_{\text{QoE}}^{i}, \ \forall i \in \mathcal{D} && \text{(Memory constraint)} \nonumber
% \end{align}

As training exposes a broader set of parallelism challenges and the same formulation naturally applies to inference (which omits backward propagation), we use training as the primary example throughout the paper. We later show that \name provides substantial benefits for inference too (\S\ref{sec:eval}).

\begin{figure}[t]
    \centering
    \includegraphics[width=0.48\textwidth]{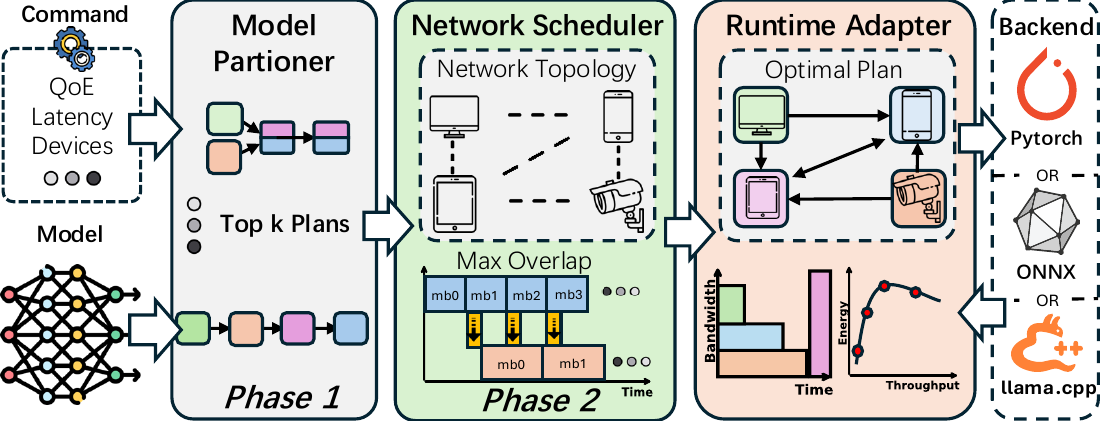}
    \caption{\name decomposes the intertwined effects of heterogeneous compute capabilities, contention-prone networks, and dynamics into three-phase optimizations. }
    \label{fig:network-scheduler-main}
\end{figure}

\paragraph{Workflow.} 
At its core, \name enables QoE-aware efficient  parallelism by decomposing the intertwined effects of heterogeneous compute capabilities, contention-prone network connections, and dynamics into three-phase optimizations, progressively pruning and refining plans to ensure decision quality while achieving second-scale decision efficiency.  
Figure~\ref{fig:network-scheduler-main} illustrates the overall workflow of \name: 
\blackcircled{1} \emph{Model Partitioner}: Upon receiving a model execution request and user-specified QoE requirements, the Model Partitioner accounts for heterogeneous device resources and generates a set of QoE-compliant candidate parallelism plans. Each plan specifies how model states (e.g., weights) are partitioned (e.g., via data, tensor, or pipeline parallelism) and placed across devices.
\blackcircled{2} \emph{Network Scheduler}: For each pre-selected QoE-compliant plan, the Network Scheduler optimizes end-to-end efficiency by maximizing communication–computation overlap under heterogeneous and contention-prone networks (e.g., by deciding when to send traffic). It then selects the plan that yields the best overall efficiency.
\blackcircled{3} \emph{Runtime Adaptor}: 
During execution, the Runtime Adaptor dynamically adjusts the chosen plan, such as by reconfiguring communication schedules or switching parallelism strategies, to maintain QoE guarantees in the presence of runtime dynamics.

% , including fluctuating resource variability, and evolving QoE targets.
\section{\name Design}
\label{sec:design}

We next introduce how \name pushes the QoE-efficiency frontier of model parallelism into an efficient three-phase  planning process without sacrificing decision quality: 

\begin{denseitemize}
\item \textbf{Phase 1 (Heter-Compute, Energy, and QoE-Aware)}: Quickly identify promising plans that satisfy QoE constraints, accounting for heterogeneous devices yet relaxing network complexity for fast pruning (\S\ref{sec:model-partitioner}).

\item \textbf{Phase 2 (Heter-Network Aware)}: Incorporate network conditions to refine Stage~1 identified QoE-compliant plans and select the global optimal (\S\ref{sec:network-scheduler}).

\item \textbf{Phase 3 (Dynamics Aware)}: Maintain QoE under runtime dynamics with fast, low-cost adjustments to model parallelism and network scheduling (\S\ref{sec:runtime-adapter}).
\end{denseitemize}

% Together, these stages enable \name to push the QoE–efficiency frontier for real-world edge AI deployments.

\subsection{Model Partitioner}
\label{sec:model-partitioner}

% The planning space for model parallelism is fundamentally intractable. Even with a fixed cross-device computation plan (\eg, a fixed assignment of model layers to devices), variations in inter-device communication scheduling can cause substantial variability in end-to-end execution latency (\S\ref{subsec:limitation}). Shared WiFi or wired links, heterogeneous topologies, and transient contention all reshape the pattern and cost of intermediate data transfers. Consequently, the planner must jointly reason about computation placement and communication behavior to optimize the end-to-end latency of a parallel execution. This joint optimization, however, yields an exponential number of feasible parallelism configurations, making exhaustive search computationally prohibitive.

The model parallelism planner must jointly consider computation placement and inter-device communication behavior (\eg, bandwidth allocation) to optimize the model's end-to-end latency. However, considering these two simultaneously induces an exponential number of feasible parallelism plans, making search computationally infeasible. 
Our key insight is to first extract a tractable subset of promising plans by temporarily relaxing network contention. 

We assume a contention-free communication environment in which every device pair $(i,j)$ can utilize its peak point-to-point bandwidth, \ie, the bandwidth attainable when the pair communicates in isolation. This assumption enlarges the feasible communication budget and therefore produces a \emph{superset} of all QoE-compliant plans: in real deployments, bandwidth can only be \emph{lower} due to contention, so any plan under the relaxed model can only become slower, not faster, under real conditions.
Under this simplified modeling, the model partitioner efficiently identifies plans that satisfy QoE constraints and are compute- and energy-efficient, which are then refined and ranked in later stages under realistic network conditions to recover the globally optimal plan (\S\ref{sec:network-scheduler}).

%\begin{figure}[t]
%    \centering
%    \begin{subfigure}{0.48\linewidth}
%        \captionsetup{labelfont=normalfont,textfont=normalfont}
%        \centering
%        \includegraphics[width=\linewidth]{fig/Dfig/model_graph.png}
%        \caption{.}
%        \label{fig:design_model_graph}
%    \end{subfigure}
%    \hfill
%    \begin{subfigure}{0.48\linewidth}
%         \captionsetup{labelfont=normalfont,textfont=normalfont}
%         \centering
%         \includegraphics[width=\linewidth]{fig/Dfig/plan_graph.png}
%         \caption{Qwen3 on multimodal tasks.}
%     \end{subfigure}
%
%    \caption{. \fan{drop it as it's part of the next DP figure.}}
%    \label{fig:design_plan_graph}
%\end{figure}

\paragraph{Planning Graph Construction.}
To capture computational dependencies of model layers, \name employs a planning graph abstraction.
%As shown in Figure \ref{fig:DPstate_annotation}, 
We represent the target model as a directed acyclic graph $G_M = (V_M, E_M)$ based on model structure, where each node denotes one or more layers. Adjacent layers whose combined size contributes less than $\Delta$ (e.g., 5\%) of total model parameters are merged into a single node. This lightweight compression preserves structural accuracy while reducing graph size, thus planning overhead.

This abstraction improves existing advances, which assume a strictly linear (chain-structured) execution order to scale planning for homogeneous clusters, based on two key observations. First, edge deployments are smaller in scale, while exhibiting greater heterogeneity in compute capability and contention-prone network conditions. Here, graph-based modeling is essential: it exposes fine-grained opportunities for parallelism and for overlapping computation with communication, yielding up to 31\% latency improvement over chain-based formulations (\S\ref{eval:ablation}).

Second, modern ML models increasingly adopt non-chain structures. For example, multimodal LLMs integrate multiple modality-specific encoders and projectors (text, image, audio) that run concurrently around a shared LLM backbone. This further motivates a graph-centric representation.

With the model represented as $G_M$, we express a hybrid parallelism plan $P$ for $G_M$ as a directed acyclic graph $G_P = (V_P, E_P)$. 
Each node $v \in V_P$ denotes a pipeline stage and is associated with a tuple $(G_v, D_v, B_v)$: 
$G_v$ is the model subgraph assigned to the stage; 
$D_v \subseteq D$ is the set of devices executing that stage. 
When $|D_v| > 1$, the stage uses \emph{data parallelism} across devices in $D_v$ on the model partition $G_v$. 
$B_v$ specifies the microbatch allocation across devices in $D_v$, where each entry $B_v[d_i]$ indicates how many samples device $d_i$ processes. 
A microbatch $B_m$ partitions the global batch $B$, enabling pipelined execution and compute–communication overlap, and satisfies $\sum_{d_i \in D_v} B_v[d_i] = B_m$ for every pipeline stage. As in prior work on distributed cloud and edge AI execution, we restrict \emph{tensor parallelism} to within a single node (e.g., edge servers with multiple GPUs) due to its high communication intensity~\cite{metis-atc24,280874}. 

Each edge $e \in E_P$ represents a dependency between \emph{pipeline parallelism} stages. 
Together, this formulation defines how the model is partitioned, how stages are mapped to heterogeneous devices, and how microbatches flow and synchronize across the distributed execution pipeline.

\paragraph{Heterogeneity- and QoE-aware Graph Partitioning.} 
\name aims to minimize total resource consumption (e.g., energy) under explicit QoE constraints by optimally partitioning the model into pipeline stages and assigning each stage to heterogeneous devices, which together form data-parallel groups. We formulate the problem as a dynamic programming (DP) search, where each DP transition (plan expansion) selects the option that minimizes the objective in Eq.~(\ref{theobjfunction}).

To make the DP tractable on graphs, we first perform a \emph{serial decomposition} of the model execution graph $G_M$ so that each decomposed component $g \in G_M$ forms an independent serial chain of nodes. We then apply DP to each component to compute its locally optimal plan. Finally, we compose all partial plans to obtain the globally optimal plan for the entire model graph $G_M$.
This DP formulation enables principled exploration of the search space over data, pipeline, and tensor parallelism, while pruning suboptimal prefixes early. 
 
% , where a plan is constructed by progressively extending a partial assignment following the topological order of the model's computation graph. Each DP transition selects the extension that maximizes the objective in Eq.~(\ref{theobjfunction}). 

% As a result, it achieves both high efficiency and strong solution quality.

\begin{figure}[t]
    \centering
    \includegraphics[width=0.9\linewidth]{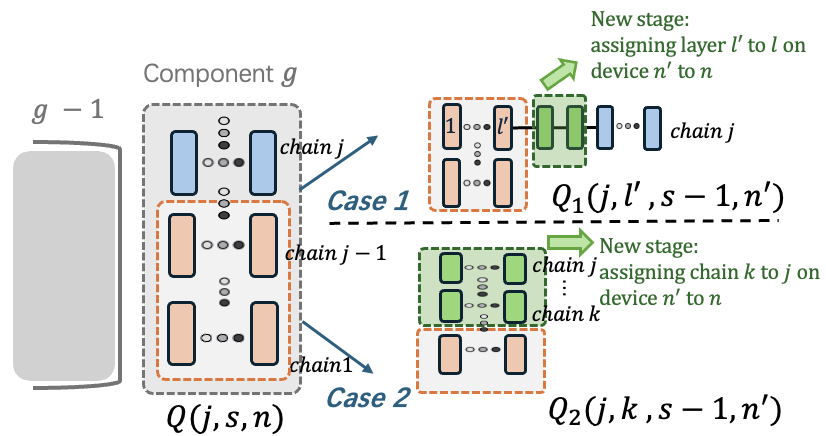}
    \caption{\name performs plan expansion to explore hybrid parallelism plans using a graph-level dynamic programming approach; each transition is optimized for the objective value.}
    \label{fig:DPstate_annotation}
\end{figure}

% We first perform a serial–then–parallel decomposition of the model graph 
% $G$, partitioning it into several serial $H$ phases. Each phase consists of multiple independent single-chain paths. For common model structures such as multi-modality models, this decomposition typically yields two phases: the first containing the individual modality-encoder chains, and the second containing the cross-modality backbone; while common transformer-based LLM yields a single phase with single chain.

As shown in Figure~\ref{fig:DPstate_annotation}, within a decomposed component $g$, the DP state is defined as 
$Q(j,s,n)$, which denotes the optimal plan for partitioning the first $j$ chains into $s$ pipeline stages and assigning these stages to the first $n$ devices. When we update $Q(j,s,n)$, we consider two cases:  
(1) the $j^{\text{th}}$ chain spans multiple pipeline stages, or  
(2) the $j^{\text{th}}$ chain is entirely contained within a single pipeline stage (possibly bundled with several preceding chains).  
We introduce two DP substates to capture these cases:
\begin{denseitemize}
    \item $Q_1(j, l, s, n)$: the best objective value obtained by assigning the first $j-1$ chains and the first $l$ layers of the $j^{\text{th}}$ chain into $s$ pipeline stages, and assigning these stages to the first $n$ devices;

    \item $Q_2(j,k, s, n)$: the best objective value obtained by assigning chains $k, k{+}1, \ldots, j$ entirely to a single pipeline stage, while partitioning the preceding $k-1$ chains into $s-1$ stages, and finally assigning all resulting stages to the first $n$ devices.
\end{denseitemize}
These states' updates are given by:

\begin{align}
\label{des:opt1.dpQ1update}
& Q_1(j,l,s,n) =
    \min_{l', n'}
    \widehat{Q}_1(j,l' \to l,\, s-1,\, n' \to n)
\\[6pt]
\label{des:opt1.dpQ2update}
& Q_2(j,k,s,n) =
    \min_{n'}
    \widehat{Q}_2(k \to j,\, s-1,\, n' \to n)
\\[6pt]
\label{des:opt1.dpQfupdate}
& Q(j,s,n) =
    \min\!\bigl(
        Q_1(j,L,s,n),\ 
        \min_{1\le k<j} Q_2(j,k,s,n)
    \bigr)
\end{align}
where $L$ is the total number of layers in the $j^{\text{th}}$ chain.

\begin{algorithm}[t]
\caption{QoE-aware Hybrid Parallelism}
\label{des:opt1.dp}
\begin{algorithmic}[1]

%\State \textbf{SET} $G_L$, D, B, M, $bandwidth$ as \textbf{Global}

%\Function{DynamicPartitioner}{}
%\State \textbf{Initialize} $top_kContainer$
%\State series decompose $G_L$ to get $H$ phases
%\For{$i$ from $1$ to $H$}
%    \State parallel decompose each subgraph of $i^{th}$ phase to individual chains
%
%    \EndFor
%\EndFor
%\State \textbf{Update} $top_kContainer$ from Eq. \ref{des:opt1.dpscope}
%\State \textbf{return} $top_kContainer$
%\EndFunction
\Function{ParallelismPlanner}{$G_M$, $D$}

% \State get device amount $N$ from device list $D$,
% \State get nodes/layers set $|V_M|$ form $G_M$
\mycommfont{// Identify Top-K computation-optimized plans (\S\ref{sec:model-partitioner})}

\State plans $\gets$ ModelPartitioner($G_M$, $D$)
\State plan\_candidates $\gets$ SelectTopK(plans)

\mycommfont{// Perform network-aware plan refinement (\S\ref{sec:network-scheduler})}
\State plan\_candidates $\gets$ NetworkScheduler(plan\_candidates)

\mycommfont{// Perform runtime plan adaptation (\S\ref{sec:runtime-adapter})}
\State hybrid\_plan $\gets$ RuntimeAdaptor(plan\_candidates)

% \For{$s$ from $1$ to $min(N,|V_M|)$}
%     \For{$g$ in $C$}
%         \State COMPONENTSEARCH($g$, $s$, $Q$)
%         \State \textbf{QoE Violation Check}\fan{let me work on this. you clean up the next DP func}
%     \EndFor 
%     Plan
% \EndFor

% \EndFor
\State \textbf{return} hybrid\_plan
\EndFunction
\\
\Function{ModelPartitioner}{$G_M$, $D$}

% \state Parse $g$, get chain amount $J$
% \state get nodes/layers amount $N_i$ for $i^{th}$ chain
\State Serial decompose $G_M$ to component set $C$

\For{$j$ from $1$ to $J$}
    \For{$n$ from $1$ to $N$}
        \State \textbf{Initialize} $Q_1(j,0,s,n)$
        \For{$l$ from $1$ to $N_j$}
            \State \textbf{Update} $Q_1(j,l,s,n)$ \label{alg:update1} 
                from Eq.~(\ref{des:opt1.dpQ1update}) as $Q_1$
        \EndFor

        \For{$k$ from $1$ to $j$}
        \State \textbf{Update} $Q_2(j,k,s,n)$ \label{alg:update2} 
                from Eq.~(\ref{des:opt1.dpQ2update}) as $Q_2$

        \State \textbf{Update} $Q(j,s,n)$ \label{alg:updatef} 
                from Eq.~(\ref{des:opt1.dpQfupdate})
    \EndFor 
\EndFor
\EndFor
\State \textbf{return} $P$ 
\EndFunction
\end{algorithmic}
\end{algorithm}

Algorithm~\ref{des:opt1.dp} illustrates our core parallelism algorithm. We process components following their topological order in the graph; for each component, the DP iteratively expands partial plans until the entire model graph $G_M$ is covered, yielding the final device-parallel partition.

For the transition in Eq.~(\ref{des:opt1.dpQ1update}) (Line~\ref{alg:update1}),  
$\widehat{Q}_1$ denotes the objective of a plan obtained by combining:
(1) an optimal plan $Q_1(j, l', s-1, n')$, which partitions the first $j-1$ chains and first $l'$ layers of $j^{th}$ chain to $s-1$ stages, and assign stages across the first $n'$ devices; 
(2) a new pipeline stage containing layers $(l'+1)$ through $l$ of chain $j$, assigned to devices $(n'+1)$ through $n$. 
The base case $Q_1(j, 0, s{-}1, n)$ naturally degenerates to $Q(j{-}1, s{-}1, n)$.

For the transition in Eq.~(\ref{des:opt1.dpQ2update}) (Line~\ref{alg:update2}),  
$\widehat{Q}_2(k \!\to\! j, s{-}1, n' \!\to\! n)$ consists of: 
(1) an optimal prefix plan $Q(k{-}1, s{-}1, n')$ that partitions the first $(k{-}1)$ chains into $(s{-}1)$ stages over $n'$ devices, and
(2) a new pipeline stage containing the entire consecutive block of chains $k$ through $j$, assigned to devices from $(n'+1)$ to $n$.

\paragraph{Achieving Load Balance for Hybrid Parallelism.}
Our DP formulation above optimizes execution latency for the given micro-batch size using data, pipeline, and tensor parallelism. However, data parallelism in training introduces an additional challenge: the micro-batch size assigned to each data-parallel replica must be balanced across devices. Otherwise, the slowest replica elongates the iteration during gradient aggregation.

To achieve load-balanced execution, we allocate micro-batches proportionally to device speeds.  
For a replica group with $x$ devices, let $N_{b,i}$ be the number of micro-batches assigned to device $i$.  
If $T_i$ denotes the per-micro-batch latency of device $i$, the execution time of that replica is approximately linear in its workload, \ie,
$f(N_{b,i},T_i) \propto N_{b,i} \cdot T_i$.  
Hence, balancing workloads reduces to assigning
\[
N_{b,i} \approx 
\frac{1/T_i}{\sum_j (1/T_j)} \cdot \frac{M}{b},
\]
where $M$ is the global batch size and $b$ is the micro-batch size.  
This simple proportional rule yields near-optimal load balance while avoiding an expensive integer optimization.

% Here, $f_{\mathcal{L}}(N_{b,i}, T_i)$ denotes the latency of executing partition $\mathcal{L}$ on device $i$ with profile $T_i$. 
% Assuming latency scales linearly with batch size and layer count, 
% $f_{\mathcal{L}}(N_{b,i}, T_i) = C \cdot N_{b,i} \cdot T_i \cdot |\mathcal{L}|$, 
% where $C$ depends on the model structure. 
% This simplifies the problem to an integer optimization, with 
% $N_{b,i}$ approximated by $\frac{R_i}{\sum_j R_j} \cdot \frac{M}{b}$, 
% where $R_i = 1 / T_i$ represents device performance.

Thus far, we have selected the parallelism plan based on ideal network conditions, where each device pair can use its peak bandwidth. In practice, however, contention---e.g., shared WiFi bandwidth or wired links---can change the end-to-end performance. Our key insight is that, even though the true optimal plan under real network conditions may not be the best plan, it would appear within the top few candidates: all plans are ranked under the same idealized network, the best real-world plan still tends to stay near the top. Therefore, at each DP transition (Line~4), \name keeps the top-$K$ most promising plans. These top-$K$ heterogeneity-aware candidates are then passed to the next phase (\S\ref{sec:network-scheduler}) for refinement under realistic network conditions.

\subsection{Network Scheduler}
\label{sec:network-scheduler}

% Thus far, the Model Partitioner (Stage~1) determines how to distribute the model across heterogeneous computational devices under an \emph{overcommitted network assumption}: every device pair is treated as having identical effective D2D bandwidth, i.e., $b_{d,d'} = b_{\text{shared}}$. This abstraction enables efficient planning and yields a \emph{superset} of all QoE-compliant parallelism plans, since reducing actual bandwidth can only degrade––never improve––the feasibility of a candidate computational partition. However, due to heterogeneous network topology and dynamic contention (e.g., shared WiFi bandwidth), some plans that appear feasible under the overcommitted-bandwidth assumption may be suboptimal or even infeasible when executed on real networks.

\begin{figure}[t]
\centering
\centering
\includegraphics[width=.98\columnwidth]{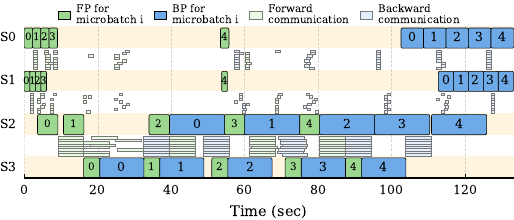}
\caption{A toy example illustrating communication and computation overlap across microbatches during model training.}
\label{fig:toy-com}
\end{figure}

We now introduce a \emph{network-aware scheduler} to refine the top candidate plans produced in Phase~1. Its goal is to maximize communication--computation overlap---as illustrated by the toy example in Figure~\ref{fig:toy-com}---across both forward/backward propagation and micro-batch execution. 
Crucially, this refinement accounts for real network conditions, including asymmetric links, bandwidth heterogeneity, irregular topologies, and runtime contention. 
The scheduler considers these factors to refine candidate plans, and ultimately selects the refined plan that achieves the best end-to-end efficiency while satisfying the QoE constraints.

\paragraph{Communication-Expanded Planning Graph.}
To account for these network effects, we extend the original planning graph $G = (V, E)$ (\S\ref{sec:model-partitioner}) into a \emph{communication-expanded planning (CEP) graph} $G' = (V', E')$. The CEP graph augments $G$ by inserting communication nodes (tasks) and corresponding dependencies, while preserving the durations and ordering of all computation nodes (tasks).

Each node $v^{\text{type}}_{i,j}$ in $G'$ represents either a computation or a communication task for the $j$-th microbatch at the $i$-th pipeline stage. The edge set $E'$ captures all execution dependencies, including inter-device transfers (\eg, sending activations or gradients between layers on different devices). Each node $i$ is annotated with a compute duration $D_i$ and a bandwidth demand $B_i$. For computation nodes, $B_i = 0$. For a communication node that transfers $T$ units of data, the bandwidth--duration product satisfies $D_i \cdot B_i = T$. This formulation enables the scheduler to flexibly trade transfer time for allocated bandwidth under contention, slowing down non-critical transfers to preserve bandwidth for latency-sensitive ones, improving overall performance.

Formally, let $I_i = [F_i, T_i]$ denote the scheduled start and finish times of node $i$, and let $R_{\text{bw}}$ denote the total available bandwidth on each link per time unit. The network-aware scheduling problem can then be formulated as the following  Linear Programming (LP) problem:

\begin{align}
\small
\min_{I} \quad & \max_i T_i && \text{(Minimize E2E latency)} \label{eq:lat-objective} \\
\text{s.t.} \quad 
& T_i = F_i + D_i && \text{(Node duration)} \nonumber \\
& T_j \leq F_i,\ \forall (j, i) \in E' && \text{(Dependency constraints)} \nonumber \\
& \sum_{i: F_i \leq t < T_i} B_i \le R_{\text{bw}},\ \forall t && \text{(Bandwidth feasibility)} \nonumber
\end{align}

This formulation jointly determines the execution order of communication tasks and their bandwidth allocations, allowing the network scheduler to minimize end-to-end latency under contention while preserving all computation tasks and dependencies produced by the Model Partitioner.  It offers two additional advantages: First, it generalizes to diverse network topologies (\eg, ring-based LANs, hierarchical wired networks, or shared WiFi) and accommodates heterogeneous bandwidth distributions.
Second, solving this LP is lightweight in edge AI settings: the number of devices, which is the size of each communication-expanded planning graph, is inherently small.  
Our evaluation across representative real-world edge deployments shows that \name computes the network-aware scheduling decisions for each candidate plan within sub-second latency (\S\ref{eval:ablation}).

% solving it is NP-hard because it corresponds to the classical NP-hard \emph{Resource-Constrained Project Scheduling Problem} (RCPSP)~\cite{}, where each task has a fixed duration and resource demand, and the objective is to compute a resource-feasible schedule with minimal makespan. As a result, invoking a full ILP solver is impractical for edge environments requiring fast responsiveness to handle dynamics: the number of computation and communication nodes, along with their dependency edges—i.e., $|V'| + |E'|$—quickly grows into the hundreds for modern ML models. Even a relatively small model such as Qwen3-0.6B contains roughly 30 Transformer layers, and the expanded graph further multiplies in size with the number of microbatches (typically 4 or more), all of which must be jointly scheduled to optimize end-to-end pipeline efficiency.

% \fan{We need a figure showing the scalability (overhead) of the full RCPSP solver.} 
% To make the approach practical for online and large-scale deployment, we next introduce two heuristic strategies that effectively reduce the planning space while retaining near-optimal scheduling efficiency.

\paragraph{Jointly Optimizing Communication and Computation.}
Our lightweight CEP design allows \name to refine multiple candidate plans efficiently (in seconds~\S\ref{eval:ablation}) and even in parallel, since plans are independent once produced by the Model Partitioner. \name retrieves the Top-$K$ candidates from Phase~1, constructs a communication-expanded planning graph for each, optimizes its schedule, and selects the plan achieving the highest overall efficiency.

However, explicitly controlling bandwidth allocation in edge environments is often difficult or infeasible (\eg, requiring privileged access to the WiFi Access Points).  
To overcome this limitation, \name transforms \emph{spatial} bandwidth sharing into \emph{temporal} sharing by splitting each communication transfer into multiple data chunks.  
Concretely, a transfer of total size $T$ is divided into $w$ subtasks, each modeled as an independent communication task with duration:
$
D_i = \frac{T}{R_{\text{bw}} \cdot w}.
$
This chunking strategy allows devices to effectively control when and how many chunks to transmit, thereby realizing the intended network scheduling behavior \emph{without} requiring direct manipulation of physical bandwidth allocation.

Combining the Phase~1 and ~2 optimizations, our evaluation shows that \name delivers \emph{near-optimal} performance while completing end-to-end planning within seconds (\S\ref{eval:e2e}).  

% \lee{I hope that we show how far our three-stage process is from an optimal solution by solving eq (1) directly since we claim near-optimality here.}

% Second, to avoid an explosion in dependency edges during graph expansion (\eg, many communication subtasks), we introduce two dummy nodes to wrap the $w$ subtasks: one before and one after the group. If the original task had $a$ successors and $b$ predecessors, directly propagating these to each subtask would introduce $(w{-}1)(a + b)$ new edges. Instead, by routing all predecessor edges to the first dummy node and all successor edges from the second dummy node, we reduce the total to just $2w$ edges, greatly improving solver scalability. 

% \fan{I don't see where our greedy algorithm is.}
% By solving Eq~(\ref{eq:objective}), we can extract the execution order of communication subtasks to overlap with computation tasks, while fully utilizing the network bandwidth. 

% \fan{we are supposed to discuss top-k plan somewhere?}
% \jl{currently,consider a 4.3 section named like "putting all together", and shold generate a highest level sudo code including profiling, do algorithm1 and then do wavefron scheduler for each candidates...}

\subsection{Runtime Adapter}
\label{sec:runtime-adapter}

At runtime, edge environments are inherently dynamic: device compute capacity and network conditions can fluctuate (e.g., due to competing foreground or background workloads), and devices may even drop out due to mobility or energy constraints. Such variability can quickly render a static parallelism plan suboptimal. 

\name incorporates a lightweight \emph{Runtime Adapter} that performs fast, on-the-fly replanning to maximize long-term efficiency gains. It maintains high efficiency and QoE by satisfying two deployment-driven objectives:  
(1) \emph{Maximized Efficiency} for interruptible workloads (e.g., training or tuning). These workloads often optimize for long-term QoE metrics, such as ``\texttt{completing a tuning job by 9~AM},'' so minor pauses are acceptable; and 
(2) \emph{Minimal Service Interruption} for continuous workloads (e.g., online serving), where QoE (and model execution) should remain uninterrupted as much as possible, even at the cost of efficiency loss.

\begin{figure}[t]
\centering
\centering
\includegraphics[width=\columnwidth]{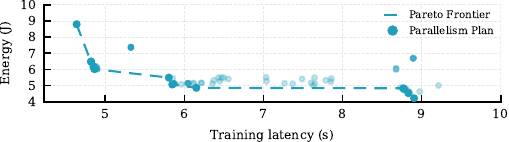}
\caption{Parallelism plans exhibit diverse latency-energy tradeoffs, allowing meeting QoEs with mixed plans. }
\label{fig:qoe-efficiency}
\end{figure}

\paragraph{Maximizing Long-term Efficiency Gains.} 
Interruptible workloads offer greater flexibility to pursue more aggressive efficiency optimization: as shown in Figure~\ref{fig:qoe-efficiency}, different parallelism plans present different QoE-efficiency tradeoffs, so we can \emph{strategically mix} multiple plans over time to maximize overall efficiency without violating long-term QoE targets. For example, suppose Plan~A would miss the deadline if used alone, but is highly energy-efficient. \name can first execute under Plan~A to harvest its energy savings, and later switch to a faster plan to ensure the job still completes before the deadline. 

% This adaptive combination yields better global efficiency while preserving the required QoE.

However, this approach introduces two key challenges:  
(1) How should \name identify the right set of plans to combine over time to maximize efficiency while still meeting long-term QoE? and 
(2) How can the system remain robust to uncertainties—such as resource fluctuations—that may later invalidate  ``fast'' plans and risk violating the QoE deadline?

To address both challenges, we introduce a \emph{uniform-progress} heuristic that amortizes the end-to-end QoE target over short decision horizons. At runtime, \name periodically assesses the remaining workload $W_{\mathrm{rem}}$ and deadline $D_{\mathrm{rem}}$, partitioning the future into horizons of length $\Delta$. For each horizon, it computes an \emph{expected progress} requirement $\mathrm{EP}_{\Delta} = (\Delta / D_{\mathrm{rem}})\cdot W_{\mathrm{rem}}$, representing the minimum useful work needed to remain on schedule. This formulation provides robustness and flexibility: if a horizon falls short due to transient slowdowns, the next horizon re-evaluates $W_{\mathrm{rem}}$ and $D_{\mathrm{rem}}$, naturally increasing $\mathrm{EP}_{\Delta}$ to compensate. Thus, deficits are automatically absorbed and corrected over time.

% , and \name can opportunistically exploit the diverse QoE–efficiency tradeoffs across plans—favoring slower, energy-efficient plans when ahead of schedule and switching to faster, costlier plans only when necessary.

Within each horizon, the Runtime Adapter selects a mixture of plans that jointly satisfy the horizon’s expected-progress target while optimizing the global QoE-aware objective introduced in Eq.~(\ref{theobjfunction}). 
% This ensures that runtime adaptation remains fully aligned with the global optimization goals established in Stages~1 and~2. 
Let $\mathcal{P}$ denote the QoE-compliant, Pareto-optimal plans generated earlier. For each plan $p \in \mathcal{P}$, we use its latency $r_p$, energy rate $e_p$, and switching overhead $d_p$ to determine the useful execution time and additional energy cost within a horizon of length $\Delta$. The decision variable $x_p \in [0,1]$ specifies the fraction of the horizon assigned to plan $p$. 
The Runtime Adapter then solves a small linear program that determines the optimal mixture:

\begin{align}
\min_{\{x_p \ge 0\}} \quad 
& \text{Obj}\bigl(x_p;\, \text{Eq.~\ref{theobjfunction}}\bigr)
    \label{eq:rt_obj_simplified}
\\[3pt]
\text{s.t.}\quad 
& \sum_{p \in \mathcal{P}} r_p\, x_p(\Delta - d_p) \ \ge\ \mathrm{EP}_{\Delta},
    \label{eq:rt_progress_simplified}
\end{align}

Constraint~(\ref{eq:rt_progress_simplified}) ensures that the horizon completes sufficient work to maintain uniform progress toward the QoE goal under useful execution time $\Delta-d_p$.

This formulation enables \name to react quickly to resource variations without endangering long-term QoE. Because the LP is extremely small, involving only a handful of plans, it can be solved in milliseconds (\S\ref{eval:e2e}).

% However, this reactive policy alone does not account for the \textit{plan-switching cost} $d$, which captures data migration, warmup, and re-scheduling overheads. Importantly, $d$ can be much higher for \textit{training} (where weights are sharded across devices) than for \textit{inference} (where models are locally cached). When $d$ becomes comparable to the expected runtime between replanning intervals, frequent plan switching may harm stability and throughput.

% To prevent oscillations, we introduce a \textbf{Plan-Stay Rule}:  
% if $\text{CP}(t) < \text{EP}(t)$ but no failure or constraint violation is observed, the current plan persists for at least one iteration unless a candidate plan offers sufficiently greater improvement. Formally, let $T_{\text{rest}}$ denote the remaining time to complete the current iteration, then define:
% \[
% T_{\text{bound}} = T_{\text{rest}} - d.
% \]
% A new plan replaces the current one only if its predicted latency $T_{\text{planlatency}} < T_{\text{bound}}$.
% This ensures that a switch occurs only when it yields net progress benefit.

\paragraph{Achieving Minimal Service Interruption Under Dynamics.} 
For continuous workloads, when runtime conditions fluctuate, \name first attempts to accommodate these changes by adjusting the network scheduling derived in Stage~2 without altering the computation plan. This strategy handles transient dynamics such as short-lived bandwidth drops or momentary device QoE downgrades. As network schedule adjustments can be computed and enforced within subseconds (\S\ref{sec:network-scheduler}), this allows \name to avoid any service stalls for the majority of runtime disturbances.

For substantial or persistent changes—such as large, sustained shifts in device capability or QoE requirements, or cases where adjusted network scheduling cannot satisfy QoE, replanning and plan switching become unavoidable for any model parallelism workflows. Fortunately, \name's planning overhead is minimal (only a few seconds), so the dominant cost is the switching of model states (e.g., loading model weights). To further reduce this cost, \name introduces two complementary optimizations once a new plan is selected based on the current resource settings: (1) \emph{Asynchronous Switching}: While the model continues executing under the current plan, \name proactively begins transferring immutable states required by the new plan. For example, model weights used in inference are immutable, allowing devices to load weights for the new placement in the background without interrupting ongoing execution; and (2) \emph{Delta Switching}: Devices transfer only the \emph{missing} states that differ between the old and new plans. For example, a device fetches weights only for layers newly assigned to it.

Together, our evaluation shows that \name can complete adaptation, including migration, within a few seconds (\S\ref{eval:ablation}).

% Dynamic adaptation at runtime introduces two central challenges: (i) \emph{when to migrate?} Frequent plan switching incurs non-trivial overhead due to model migration and scheduling disruptions. Thus, \name must determine \emph{when} adaptation is worth the cost; (ii) \emph{How to migrate?} Once a plan switch is warranted, \name must efficiently select the next best plan to transition to, ideally without full reprofiling or large data movement.

% \paragraph{Complexity Analysis.} 
% \fan{1. quickly summarize how these three stages coordinate; 2. for each, analyze the complexity like O(N).}
% By combining timeout reprofiling with dynamic planning policy, \name is capable of gracefully handling device resource fluctuations while still respecting user-defined latency goals. It is important to note that \name does not guarantee meeting the QoE latency deadline, denoted as $R$ before, as future device availability cannot be precisely predicted. However, when latency targets are reasonably achievable, \name aims to meet them and meanwhile prioritizes energy efficiency wherever possible.
%\input{tex/design}
\section{Implementation}
\label{sec:imple}

We implemented a prototype of \name in roughly 4{,}000 lines of code, supporting distributed edge AI training and inference atop ONNX Runtime~\cite{onnx-github} and PyTorch. 

\name's execution backend enables hybrid model parallelism across heterogeneous devices. The system is implemented in Python and builds on PyTorch PiPPy and Distributed Data Parallel (DDP). Our design composes existing single-parallelism libraries while using PyTorch's low-level communication primitives to support efficient data exchange. A key requirement of \name's phase-two network scheduler is the ability to fragment communication transfers into fine-grained sub-transfers to enable network scheduling.

\name designates the most capable device as a coordinator, which issues periodic heartbeats to track device health and resource status. When fluctuations remain small (within 10\%), the runtime adapter invokes only the phase-two network optimization, avoiding the overhead of a full re-partition. The heartbeat mechanism also functions as a lightweight failure detector: if a device or the coordinator becomes unresponsive, the system initiates a consensus-based recovery protocol that triggers re-planning and ensures continued progress.

% \paragraph{Simulator}
% \name utilizes a custom profiling module to acquire offline configurations of participating devices and the network topology. Leveraging this precise profiling data, we implemented an accurate simulator capable of estimating latency and energy consumption for both \name and baseline methods. The simulator models complex network flows and supports common network control mechanisms, such as TCP protocols. Furthermore, it estimates the planning time overhead for \name and comparative baselines. In stable environments, the simulator's latency estimates remain within a 10\% margin of error compared to actual deployment.
\section{Evaluations}
\label{sec:eval}

We evaluate \name across four realistic edge AI deployments and diverse model families. Our key findings are:

\begin{denseitemize}
    \item \name discovers better model parallelism plans, achieving 1.1-- 6.3$\times$ faster training and inference (\S\ref{eval:e2e});
    
    \item \name reduces energy consumption by 20.7--82\% while fully meeting QoE requirements (\S\ref{eval:e2e});
    
    \item \name consistently outperforms existing advances across diverse devices, networks, and workload scales (\S\ref{eval:ablation}).
\end{denseitemize}

\subsection{Experiment Setup}
\label{eval:setup}

\begin{table}[t]
\centering
\label{eval:model_table_swapped}
\small
\begin{tabular}{l c c c c}
\toprule
\textbf{} & Bert & Qwen3-0.6B & Qwen3-1.7B & Qwen-Omni \\
\midrule
Model Size & 0.1B & 0.6B & 1.7B & 6B \\
\bottomrule
\end{tabular}
\caption{Our evaluations span model training and inference of multiple sizes, including both LLMs and MLLMs.}
\label{tab:models}
\end{table}

% \begin{table}[t]
% \small
% \label{eval:dataset_table}
% \scriptsize
% \setlength{\tabcolsep}{4pt}
% \resizebox{\columnwidth}{!}{
% \begin{tabular}{l l r r}
% \toprule
% \textbf{Task} & \textbf{Dataset} & \textbf{Example Size} & \textbf{Request Size} \\
% \midrule
% \multirow{3}{*}{Multi-modality Q\&A} 
%     & Alpaca\cite{alpaca} & 32,392 & 1,800 \\
%     & Inmsys-chat-1m\cite{inmsys} & 273,043 & 15,170 \\
%     & OpenOrca\cite{openorca} & 774,285 & 43,016 \\
% \midrule
% \multirow{2}{*}{Question Q\&A}
%     & MS MARCO\cite{msmarco} & 808,731 & 101,092 \\
%     & Natural Questions\cite{nq} & 300,000 & 7,830 \\

% \bottomrule
% \end{tabular}
% }
% \caption{\textbf{Datasets including text task and MM task}}
% \end{table}

\paragraph{Models and Datasets.}
As shown in Table~\ref{tab:models}, our evaluation spans modern AI models of multiple sizes, 
including both LLMs and multimodal LLMs with audio and image encoders. 
For text-based models (e.g., BERT and Qwen3-1.7B), we use the real-world LMSys-Chat dataset~\cite{lmsys-chat} for both training and inference experiments. 
Note that \name operates strictly at the systems level and does not alter model quality. 
For multimodal workloads, we evaluate the Qwen2.5-Omni-7B model on the MMMU-Pro benchmark~\cite{mmmu-arxiv}, covering diverse domains such as digital farming and medical AI, as well as on AudioSet~\cite{gemmeke2017audioset}, a large-scale dataset containing both visual and auditory inputs.

% For audio modality part of MLLM taskset, we synthesized audio input by our own. 

\paragraph{Edge Environment Setup.}

\begin{table}[t]
\label{eval:devices_info_table}
\centering
\small
\begin{tabular}{l l c}
\toprule
\textbf{Edge Device} & \textbf{Accelerator} & \textbf{Memory} \\
\midrule
Samsung Galaxy S25  & Snapdragon 8 Elite & 12GB \\
Xiaomi 15    & Snapdragon 8 Elite  & 12GB \\
MediaTek Genio 520      & MediaTek 8th Gen NPU   & 16GB \\
MediaTek Genio 720      & MediaTek 8th Gen NPU   & 16GB \\
Alienware 16 Laptop      & RTX 4050   & 6GB \\
Lenovo Laptop      & RTX 4060   & 8GB \\
Edge Server 1     & NVIDIA V100   & 16GB \\
Edge Server 2     & NVIDIA A40   & 16GB \\
\bottomrule
\end{tabular}

\caption{Specifications of edge devices used in experiments.}
\label{tab:devices_info_table}
\end{table}

\begin{table}[t]
\label{eval:setting_table}
\centering
\small
\begin{tabular}{l l l}
\toprule
\textbf{Setting} & \textbf{Devices} & \textbf{Network Topo}\\
\midrule

Smart home 1 & 2×4060ti, 3×4050 & 900 Mbps (WiFi) \\
Smart home 2 & 2×4050, 2×Mi15, S25 & 600 Mbps (WiFi) \\
Traffic monitor & 2×M720, 2×M520 & 200 Mbps (Wired Ring) \\
Edge Cluster & 2×A40, 2×V100 & 4 Gbps (Wired Ring) \\

\bottomrule
\end{tabular}
\caption{Evaluations span four representative edge  settings.}
\label{tab:setting_table}
\end{table}

\begin{figure*}[t]
\centering
\includegraphics[width=0.9\textwidth]{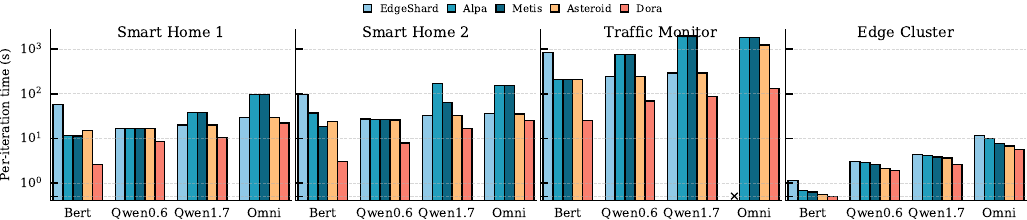}
\caption{\name achieves $1.1$--$6.3\times$ better training latency across settings and models.}
\label{fig:train_latency}
% \end{subfigure}
\end{figure*}

\begin{figure}[t]
\centering
\centering
\includegraphics[width=\columnwidth]{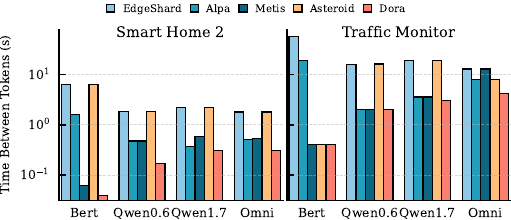}
\caption{\name achieves better serving latency by $1.2$--$2.8\times $.}
\label{fig:infer_latency}
\end{figure}

We evaluate \name using six heterogeneous edge devices and two commodity edge servers (Table \ref{tab:devices_info_table}). Our evaluations span four representative edge environments (Table \ref{tab:setting_table}): Smart Home 1 and Smart Home 2 model typical households equipped with laptops, desktops, and smartphones, but with different WiFi capabilities and in-home bandwidth constraints. Traffic Monitoring represents a deployment where smart cameras at an intersection collaboratively analyze video and audio streams; these devices are connected through a ring-style topology commonly found in multi-camera sensing systems~\cite{ring-topo}. Finally, Edge-Server captures small lab or enterprise clusters consisting of GPUs connected via high-speed, low-contention links.

\paragraph{Baselines.}
We compare \name against multiple state-of-the-art model parallelism planners:

\begin{denseitemize}

\item \emph{Asteroid}~\cite{ye_2024_asteroid}: a model parallelism planner for edge AI \emph{training}, yet overlooks network contention and topology.

\item \emph{EdgeShard}~\cite{zhang2024edgeshard}: a distributed training framework using pipeline parallelism. 

% Although it considers communication latency, it employs simplified abstractions for bandwidth constraints.

\item \emph{Alpa}~\cite{280874}: an advanced parallelism planner combining data, pipeline, tensor parallelisms for cloud AI training.

\item \emph{Metis}~\cite{metis-atc24}: a device heterogeneity-aware planner for cloud AI training.

\end{denseitemize}

\paragraph{Metrics.}
\name aims to improve the following metrics for both training and inference workloads:
\begin{denseitemize}
\item \emph{Latency}: per-iteration duration for model training and tuning jobs, and serving latency for inference workloads (e.g., per-token generation latency in LLMs).

\item \emph{Energy}: 
the amount of energy consumption in meeting the QoE requirements.

\item \emph{Responsiveness}: the reaction time to decide a plan. 
\end{denseitemize}

We report the mean values over five runs per experiment.

% ============================
% Row 3 : b d (half + half)
% ============================
%\begin{subfigure}{0.48\textwidth}
%    \centering
%    \includegraphics[width=\textwidth]{fig/Efig/inference.pdf}
%    \caption{Inference serving latency}
%    \label{fig:infer_latency}
%\end{subfigure}
%\hfill
%\begin{subfigure}{0.48\textwidth}
%    \centering
%    \includegraphics[width=\textwidth]{fig/Efig/inference_energy.pdf}
%    \caption{Inference energy consumption}
%    \label{fig:infer_energy}
%\end{subfigure}

% \caption{\name achieves better (a) training latency and (b) energy savings in maintaining QoE requirements.}
% \label{fig:all_graphs}
% \end{figure*}

\subsection{End-to-end Performance}
\label{eval:e2e}

% \begin{figure*}[t]
% \centering

% % ============================
% % Row 1
% % ============================
% \begin{subfigure}{0.66\textwidth}
%     \centering
%     \includegraphics[width=\textwidth]{fig/Efig/training.pdf}
%     \caption{Training iteration latency}
%     \label{fig:train_latency}
% \end{subfigure}
% \hfill
% \begin{subfigure}{0.33\textwidth}
%     \centering
%     \includegraphics[width=\textwidth]{fig/Efig/inference.pdf}
%     \caption{Inference serving latency}
%     \label{fig:infer_latency}
% \end{subfigure}

% \vspace{1em} % Space between rows

% % ============================
% % Row 2
% % ============================
% \begin{subfigure}{0.66\textwidth}
%     \centering
%     \includegraphics[width=\textwidth]{fig/Efig/training_energy.pdf}
%     \caption{Training energy consumption}
%     \label{fig:train_energy}
% \end{subfigure}
% \hfill
% \begin{subfigure}{0.33\textwidth}
%     \centering
%     \includegraphics[width=\textwidth]{fig/Efig/inference_energy.pdf}
%     \caption{Inference energy consumption}
%     \label{fig:infer_energy}
% \end{subfigure}

% \caption{End-to-end evaluations under different settings.}
% \label{fig:all_graphs}
% \end{figure*}

\paragraph{\name achieves superior model execution speed for both edge AI training and inference.}
Figure~\ref{fig:train_latency} reports training latency across models and environments; Figure~\ref{fig:infer_latency} shows the corresponding inference performance. Under the memory-constrained Traffic Monitor setting, EdgeShard fails to produce a valid plan: its four-stage pipeline overloads the first stage with activation memory, leading to repeated OOM failures. Across all environments, \name delivers consistently faster training: in the Edge Cluster setting, it outperforms the best baseline by at least $1.1\times$, and in the network-contentious Smart Home 2, it achieves $6.3\times$ faster training. Similarly, \name achieves 1.2--2.8$\times$ faster inference.

A notable pattern emerges in Qwen-0.6B training under Smart Home 1 and Smart Home 2: all baseline methods converge to nearly identical latencies. This plateau arises because communication over the shared network becomes the dominant bottleneck. Existing baselines either optimize only computation or simplify the network into idealized D2D links, ignoring bandwidth contention. As a result, they collapse to the same suboptimal plan once communication saturates. In contrast, \name explicitly models shared-link contention and orchestrates communication to mitigate it, yielding sustained speedups of $1.9\times$ and $3.3\times$ over the best baseline in these challenging settings.

\begin{figure}[t]
    \centering
    \includegraphics[width=\columnwidth]{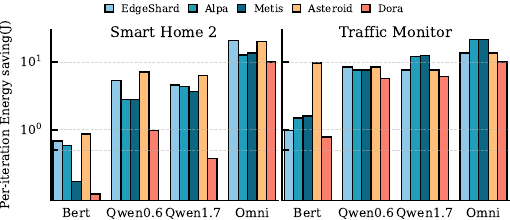}
    \caption{\name reduces energy consumption by 15\%--82\%.}
    \label{fig:infer_energy}
\label{fig:all_graphs}
\end{figure}

\begin{figure*}[t]
\centering
\includegraphics[width=0.9\textwidth]{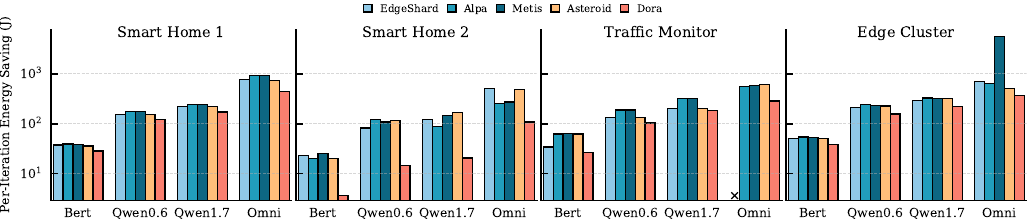}
\caption{\name reduces energy consumption by 10\%--82\% in meeting training QoE requirements.}
\vspace{.1cm}
\label{fig:train_energy}
\end{figure*}

\paragraph{\name saves energy for edge AI training and inference while meeting QoE requirements.}
Figure~\ref{fig:train_energy} shows the aggregate energy consumption of participating devices for achieving the target QoE training iteration speed, and Figure~\ref{fig:infer_energy} shows the same for inference. 
We define QoE constraint as $T_{QoE}$ as 0.8$\times$ (training and inference) latency performance of the best baseline and performed additional ablation studies on varying QoE tightness later (\S\ref{eval:ablation}). In scenarios with low device heterogeneity, such as \textit{Traffic Monitor} and \textit{Edge Cluster}, \name achieves an energy reduction of over 15\% while ensuring the QoE requirements. In the \textit{Smart Home 2} setting, which features higher heterogeneity, \name achieves substantial energy savings ranging from 57.5\% to 82.2\%. 
Similarly, across all setting \name achieves 15\% to 82\% energy savings on inference.

%\fan{is it about training or inference? we'd better to have some inference summarization.}

% We performed further ablation studies on the energy-QoE constraint trade-offs ($\lambda$) in Section~\ref{eval:ablation}.

% These results highlight that \name excels in common edge serving scenarios, effectively minimizing energy consumption while maintaining the QoE required to ensure a smooth user experience.

\begin{figure}[t]
\begin{minipage}{0.48\columnwidth}
    \centering
    \includegraphics[width=\linewidth]{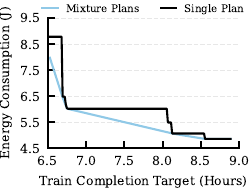}
    \caption{The Runtime Adapter attains better efficiency through mixing plans.}
    \label{fig:long-term-cost}
\end{minipage}
\hfill
\begin{minipage}{0.48\columnwidth}
    \centering
    \includegraphics[width=\linewidth]{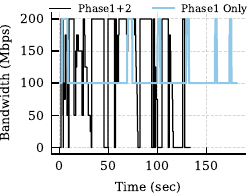}
    \caption{Network scheduler achieves efficient and flexible searching.}
    \label{fig:gantt_util}
\end{minipage}
\end{figure}

\paragraph{\name enables higher long-term effectiveness via multi-plan orchestration.}
We next evaluate \name under long-running workloads. 
Figure~\ref{fig:long-term-cost} reports the total energy consumption for completing a 6{,}000-iteration tuning job in the \textit{Smart Home 2} setting. 
We compare \name's multi-plan orchestration (\S\ref{sec:runtime-adapter}) against the best single-plan alternative, ensuring both satisfy varying job-deadline requirements from 6.5 to 9 hours.
Across all QoE regimes, \name consistently achieves a substantially lower objective, improving energy efficiency by up to 31.8\% (e.g., when the deadline requirement is 6.7 hours). 
This demonstrates that adapting the execution plan over time, rather than committing to a fixed one, yields significantly better long-term effectiveness.

\paragraph{\name is scalable and more responsive than existing approaches.}
We evaluate \name's planning overhead against baseline methods. As shown in Table~\ref{tab:planning_cost}, the phase-one model partitioner achieves subsecond planning, giving \name the flexibility to prioritize either rapid adaptation or deeper search for improved plan quality. Moreover, \name's contention-aware network scheduler exposes a tunable search space that allows the system to trade minimal accuracy loss for substantial responsiveness. As illustrated in Figure~\ref{fig:gantt_util}, this enables \name to rapidly produce new schedules under bandwidth or compute fluctuations, achieving the responsiveness needed to sustain QoE.

\begin{table}[t]
\centering
\small   % slightly smaller but still readable
\setlength{\tabcolsep}{3pt} % reduce horizontal padding
\begin{tabular}{c|ccc|ccc}
\toprule
& \multicolumn{3}{c|}{\textbf{Smart Home 2}} & \multicolumn{3}{c}{\textbf{Traffic Monitor}} \\
& Metis & Asteroid & \name & Metis & Asteroid & Dora \\
\midrule
Bert & 0.32  & 0.17 & 0.12 
      & 0.28 & 0.13  & 0.11 \\
      
Qwen-1.7B & 0.85  & 0.50  & 0.20 
      & 0.78 & 0.39 & 0.17 \\
      
Omni & 1.383  & 1.14  & 0.79 
      & 1.25 & 1.02  &  0.72 \\
\bottomrule
\end{tabular}
\caption{\name enables faster planning (sec) for edge AI.}
\label{tab:planning_cost}
\end{table}

\subsection{Ablation Studies}
\label{eval:ablation}

\begin{figure}[t]
    \centering
    \begin{subfigure}{0.49\linewidth}
        \captionsetup{labelfont=normalfont,textfont=normalfont}
        \centering
        \includegraphics[width=\linewidth]{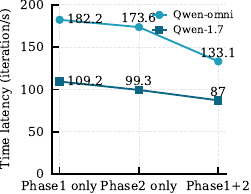}
        \caption{Training speed.}
        \label{fig:stagebreakdown_t}
    \end{subfigure}
    \hfill
    \begin{subfigure}{0.49\linewidth}
         \captionsetup{labelfont=normalfont,textfont=normalfont}
         \centering
         \includegraphics[width=\linewidth]{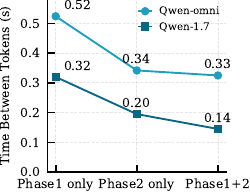}
         \caption{Inference speed.}
         \label{fig:stagebreakdwon_i}
     \end{subfigure}

    \caption{Performance breakdown of \name.}
    \label{fig:stagebreakdown}
\end{figure}

\paragraph{Performance breakdown by components.}
Figure~\ref{fig:stagebreakdown} quantifies the contributions of \name's two-phase planner to end-to-end latency. To isolate each design, we evaluate two variants: (1) \emph{Phase~1 only}, which applies \name's model partitioner without network-aware refinement; and (2) \emph{Phase~2 only}, which uses an even partitioning strategy like EdgeShard~\cite{zhang2024edgeshard} and applies \name's network-aware scheduler. 

For Qwen-Omni training, \name reduces latency by $23\%$ through Phase~2 and by $26\%$ through Phase~1. For inference, where computation dominates, Phase~1 provides up to $37\%$ latency reduction, while Phase~2 still delivers meaningful benefits---saving $25\%$ for Qwen-1.7B inference. Overall, the two phases contribute complementary optimizations, jointly enabling \name's end-to-end performance gains.

\paragraph{Performance breakdown by Time.}
Figure~\ref{fig:gantt_util} reports the network bandwidth utilization over a training iteration in the \textit{Traffic Monitor} setting, where we again break down the \name performance into two phases. We notice that with Phase 2 (network scheduler), \name achieves faster iteration by substantially improving the network resource utilization. 

% \jl{modify, we don't have figure reports execution breakdown anymore}\fan{report the util line figure}
% Figure~\ref{fig:gantt_flow} reports the end-to-end execution breakdown for Qwen2.5-Omni (MLLM) training under the \textit{Traffic Monitor} setup. \name’s plan assigns the image encoder to a Genio~520, the audio encoder to a Genio~720, and arranges the remaining devices into a two-stage pipeline for the LLM backbone. Figure~\ref{fig:gantt_util} compares bandwidth utilization for the \textit{Phase~1 only} variant versus the full \name plan. The Gantt chart shows that \name strategically interleaves small transfers with large flows, keeping the network continuously busy. Over a single iteration, \name achieves a $1.7\times$ higher average bandwidth utilization and nearly saturates the link for most of the execution, while incurring negligible scheduling overhead.

% \subsection{Ablation Studies}
% \label{eval:ablation}

% \paragraph{\name Performance across Scales.}
% here we need a graph for the RCPSP efficiency?
% \jl{not sure this section: similar to Dora is more scalable and responsive than existing
% approaches. in 6.2}\fan{consider a senario: x-axis: \# of edge devices (2, 4, 8, 16, 32) (same WiFi setting like WiFi 1 Gbps in a CS building); y-axis: the latency performance (\name, baseline.)}

\begin{figure}[t]
    \centering
    \begin{subfigure}{0.48\linewidth}
         \captionsetup{labelfont=normalfont,textfont=normalfont}
         \centering
         \includegraphics[width=\linewidth]{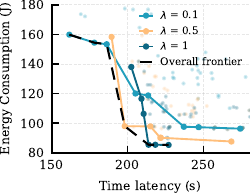}
         \caption{Training}
         \label{fig:frontier_traffic_1.7i}
     \end{subfigure}
     \hfill 
        \begin{subfigure}{0.48\linewidth}
        \captionsetup{labelfont=normalfont,textfont=normalfont}
        \centering
        \includegraphics[width=\linewidth]{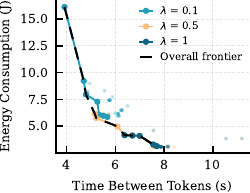}
        \caption{Inference}
        \label{fig:frontier_traffic_1.7t}
    \end{subfigure}

    \caption{\name discovers Pareto-optimal plans for different latency-energy tradeoffs ($\lambda$).} 
    \label{fig:frontier_traffic}
\end{figure}

\paragraph{Impact of QoE--Efficiency Tradeoffs ($\lambda$).}
We next examine how the trade-off parameter $\lambda$ in the objective Eq.~\ref{theobjfunction} influences the balance between energy consumption and time latency in the generated plans. Using  \textit{Traffic Monitor} setup we vary $\lambda$ from $0.1$ to $1.0$ in increments of $0.2$ for both the Qwen-1.7B model, and analyze the resulting movement of the Pareto frontier. As shown in Figure~\ref{fig:frontier_traffic}, increasing $\lambda$ shifts the frontier toward the lower-right region, reflecting a stronger preference for energy savings over latency and demonstrating the effectiveness of $\lambda$ in steering the planner's objectives. 

Notably, \name consistently produces a rich set of high-quality candidate plans along a concave frontier. Such a concave structure provides users with a flexible and well-covered spectrum of energy–latency trade-offs, and it expands the solution space available to the Runtime Adaptor for mixing plans to pursue global optimality. This behavior highlights both the efficiency of \name's search procedure and the high quality of the resulting solutions.

\paragraph{Impact of Runtime Dynamics.}
To evaluate \name's responsiveness under runtime dynamics, we run model inference in the \textit{Smart Home~2} environment using the Qwen-1.7B model. During execution, we deliberately induce interference by first downloading videos on the 4060 desktop and later rendering and watching them, thereby introducing network and compute contention to the background inference workload.

\begin{figure}[t]
\begin{minipage}{0.48\columnwidth}
    \centering
    \includegraphics[width=\linewidth]{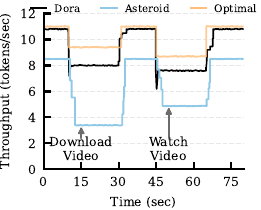}
    \caption{\name efficiently reacts to dynamics, achieving close-to-optimal performance.}
    \label{fig:dynamic-planning}
\end{minipage}
\hfill
\begin{minipage}{0.48\columnwidth}
    \centering
    \includegraphics[width=\linewidth]{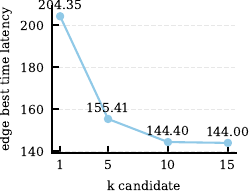}
    \caption{Impact of Top-K plans.}
    \label{fig:impact-k}
\end{minipage}
\end{figure}

Figure~\ref{fig:dynamic-planning} compares the performance of Asteroid, \name, and an \emph{optimal oracle} that instantaneously switches to the best plan upon any resource change with zero overhead. In contrast to this idealized baseline, \name leverages its Phase~2 network-aware scheduler to rapidly refine communication schedules in response to transient fluctuations. Because these adjustments avoid model-state migration, the overhead remains minimal. We observe that \name consistently tracks the optimal oracle closely and reacts within subseconds, maintaining QoE under dynamic conditions.

% \begin{figure}[t]
% \centering
% \centering
% \includegraphics[width=.9\columnwidth]{fig/Efig/graph4c.pdf}
% \caption{\name efficiently reacts to dynamics with network scheduler, while achieving close-to-optimal performance.}
% \label{fig:dynamic-planning}
% \end{figure}

\paragraph{Impact of Top-K Candidates.} 
We next ablate the number of Top-K plans generated by Phase~1 and passed to Phase~2. As anticipated in Figure~\ref{fig:impact-k}, \name achieves near-optimal performance even with a small number of candidates, validating the hypothesis that the overall best plan is typically included among the top candidates identified in Phase~1.
\section{Related Work}
\label{sec:related}

\paragraph{Edge AI Training and Inference.}
The growing adoption of edge AI accelerators has enabled deployments that improve efficiency (e.g., reduced Internet traffic~\cite{jellybean-vldb23}), accuracy (e.g., personalized models), and privacy (e.g., in-situ learning). Applications span real-world assistive agents~\cite{assistai-asset25}, digital agriculture~\cite{aifarming-2025}, and traffic monitoring~\cite{realtimeai-mobisys25}, motivating advances in quantization~\cite{qat-mobicom25}, sparsification~\cite{lte-nips24}, and lightweight model design~\cite{fedtrans-mlsys24}. However, these efforts largely optimize computation on a single device, leaving open the challenge of coordinating complex execution across heterogeneous and bandwidth-constrained edge environments.

\paragraph{Hybrid Parallelism.}
Most model-parallel systems target cloud environments. Megatron~\cite{megatron-sc21} introduces pipeline and tensor parallelism and configures them across GPUs using simple heuristics (e.g., prioritizing pipeline over data parallelism). Alpa~\cite{280874} automates data, tensor, and pipeline parallelism across homogeneous cloud GPUs, while Metis~\cite{metis-atc24} extends this automation to load-balance heterogeneous accelerators. Sailor~\cite{sailor-sosp25} jointly selects GPU placements and parallelism strategies across cloud datacenters for cost efficiency. 
% Asteroid~\cite{ye_2024_asteroid} shifts the focus to edge environments, aiming to maximize training throughput.
In contrast, \name bridges application-level QoE requirements with device-level parallel execution, explicitly modeling shared-network contention and coordinating multiple plans over time.

\paragraph{QoE-aware Serving.}
Recent systems, such as Andes~\cite{andes-arxiv24},  Tempo~\cite{tempo-arxiv25}, and MoonCake~\cite{mooncake-fast25}, have aligned LLM request serving with application-level QoE requirements by scheduling requests based on priority or slack to deadlines. While conceptually related, these systems focus on request-level scheduling in cloud datacenters and for model inference only. \name addresses a distinct challenge: QoE-aware parallelism for both model training and serving in heterogeneous, contention-prone, and dynamically evolving edge environments. As such, \name is complementary to these efforts.

\section{Conclusion}

This paper presents \name, a hybrid-parallelism planner that partitions and orchestrates model execution across heterogeneous devices for both edge training and inference. \name combines a three-phase workflow: (i) heterogeneity-aware model partitioning, (ii) network-aware scheduling to mitigate communication contention, and (iii) a runtime adaptor that mixes plans to maximize long-term efficiency under dynamics. Our evaluation across diverse models, environments, and workloads shows that \name delivers substantial latency reductions and energy savings under QoE constraints.

% \section{Citations and Bibliographies Notes}

\bibliographystyle{ACM-Reference-Format}
\bibliography{citations}
\newpage

\appendix

\section{Model Profiling}

\begin{algorithm}[h]
\caption{Efficient Plan Profiler}
\label{des:opt1.2ph}
\begin{algorithmic}[1]
\Function{StartPhaseTimeEst}{$\mathcal{P}, BList$,d}
\State S = ($2*Len(\mathcal{P})$-1) ; //get total amount of steps
\State \textbf{Initialize} $CritiPathT$;
\For{$p$ from $d$ \textbf{to} $S$}	//iterate all possible peak steps
	\State \textbf{Initialize} $CurrPathT$;
	\For{$i$ from $0$ \textbf{to} $p$}    //Necessary tasks on the critical path
		\State $CurrPathT += Bf_i$		
	\EndFor
	\State $CurrPathT += (S-p)*\max_{i \in [0, p]} Bf_i$
	
	\For{$i$ from $p$ \textbf{down to} $d+1$}    //Add rest backward dependency tasks
		\State $CurrPathT += Bb_i$
	\EndFor
	\If{$CurrPathT>CritiPathT$}
  		\State $CritiPathT = CurrPathT$
	\EndIf		
\EndFor
\State \textbf{Return} $CritiPathT$
\EndFunction

\State

\Function{EndPhaseTimeEst}{$\mathcal{P}, BList$,d}
\State S = ($2*Len(\mathcal{P})$-1) ;
\State \textbf{Initialize} $CritiPathTList$;
\For {$s$ from $0$ \textbf{to} $S-1$}     // calculate end phase time latency for each step
\State \textbf{Initialize} $CritiPathT$;
\For{$p$ from $max(s,d)$ \textbf{to} $S$}	//iterate all possible peak steps
	\State \textbf{Initialize} $CurrPathT$;
	\For{$i$ from $0$ \textbf{to} $p$}    //Necessary tasks on the critical path
		\State $CurrPathT += Bb_i$		
	\EndFor
	\State $CurrPathT += (S-p)*\max_{i \in [0, p]} Bb_i$
	
	\For{$i$ from $p$ \textbf{down to} $d+1$}    //Add rest backward dependency tasks
		\State $CurrPathT += Bf_i$
	\EndFor
	\If{$CurrPathT>CritiPathT$}
  		\State $CritiPathT = CurrPathT$
	\EndIf		
\EndFor
\State $CritiPathTList[s] = CritiPathT$
\EndFor
\State \textbf{Return} $CritiPathTList$
\EndFunction
\end{algorithmic}
\end{algorithm}

\begin{algorithm}
\caption{Plan Estimator}
\label{des:opt1.PlanE}
\begin{algorithmic}[1]
\Function{Est}{$\mathcal{P}$}

\State \textbf{Initialize} $BList$; //Generate time and energy cost for each task
\State \textbf{Initialize} $EndList$' //store the end time for each step

\State S = ($2*Len(\mathcal{P})$-1) ; //get total amount of steps

\State d = $\arg\max_{i \in \{1, \dots, S\}} (Bf_i.t+Bb_i.t)$; //get bottleneck step index

\State $T_1$ = StartPhaseTimeEst($\mathcal{P}, BList$, d)
\State $T_2$ = $(M-S+d)*(Bf_d.t+Bb_d.t)$
\State $T_{3,List}$ = EndPhaseTimeEst($\mathcal{P}, BList$, d)
\State \textbf{Initialize} $T_{latency}$; //assign a big enough initial val
\For{$i$ from $0$ \textbf{to} $S$}
	\State $T = T_1+ T_2+T_{3,List}[i]$;
	\If{$i\%2 == 0$ \textbf{and} $\mathcal{P}[i//2]$ is a device DP group}
		\State //check if the step has gathering task
  		\State $T = T+ T_{gathering}$;	//follow gathering time estimator
	\EndIf	
	\If{$T>T_{latency}$}
  		\State $T_{latency} = T$
	\EndIf	
\EndFor

\State //sum up all tasks' energy consumption
\State $E_{consumption}$ = $\sum\limits_{i=0}^{S} M*(Bf_i.e+Bb_i.e)+Bg_i.e$;	
\State \textbf{Return} $T_{latency} + \alpha*E_{consumption}$
\EndFunction

\end{algorithmic}
\end{algorithm}

\end{document}